\RequirePackage{amsmath}
\documentclass[12pt]{iopart}
\usepackage[utf8]{inputenc}
\usepackage{amsmath}
\usepackage{amsfonts}
\usepackage{amssymb}
\usepackage{graphicx}
\usepackage{hyperref}
\usepackage{bbold, bm}
\usepackage{siunitx}
\usepackage{cite}
\newcommand{\ii}{\mathrm{i}\,}
\usepackage{comment}
\usepackage{soul}
 
\usepackage{color}
\makeatletter
\renewcommand*{\@textcolor}[3]{%
  \protect\leavevmode
  \begingroup
    \color#1{#2}#3%
  \endgroup
}
\makeatother

\newcommand{\comm}[1]{}
\newcommand{\revision}[1]{\textcolor{black}{{#1}}}

\begin{document}

\title{Electron scattering by magnetic impurity in Weyl semimetals}

\author{\'{A}lvaro D\'{\i}az-Fern\'{a}ndez$^{1, *}$, Francisco Dom\'{\i}nguez-Adame$^{2}$ and Oscar de Abril$^{1}$}
\ead{$^*$ alvaro.diaz@upm.es}

\address{$^{1}$ Departamento de Estructuras y F\'{\i}sica de Edificaci\'{o}n, Universidad Polit\'{e}cnica de Madrid, E--28031 Madrid, Spain}
\address{$^{2}$ GISC, Departamento de F\'{\i}sica de Materiales, Universidad Complutense, E--28040 Madrid, Spain}

\address{$^{*}$ Author to whom any correspondence should be addressed.}


\vspace{2pc}

\date{\today}

\begin{abstract}

Weyl semimetals are prominent examples of topologically protected quantum matter. These materials are the three-dimensional counterparts of graphene and great efforts are being devoted to achieve a thorough understanding of their fundamental physics. In this work, we aim at contributing to this end by discussing the effect of a single magnetic impurity in Weyl semimetals as a first step towards considering a larger number of point-like impurities. We find that resonances appear in the local density of states with a Friedel-like behaviour, oscillating as a function of distance. By studying the spin-resolved local density of states, we can observe non-trivial and anisotropic spin textures where the spin components perpendicular to the spin of the impurity wind around the latter, until the spin becomes completely parallel to the impurity right at the interface. Friedel oscillations also play a relevant role in the form of the spin textures, forming an oscillatory pattern. We believe our results can pave the way to further studies which consider the presence of a large number of random magnetic impurities.

\end{abstract}

\submitto{\NJP}

\maketitle

\section{Introduction}

In 1929, Hermann Weyl proposed an equation which would describe the behaviour of massless relativistic particles with well-defined chirality~\cite{Weyl29,Pal11}. So far, no experimental evidence has been found to confirm the existence of elementary particles described by the Weyl equation. However, Weyl fermions have been theoretically predicted to appear as quasiparticles in solid-state settings where two energy bands meet at isolated points~\cite{Hosur13,Yan17,Shuo17,Armitage18}. These systems where Weyl quasiparticles arise are therefore dubbed Weyl semimetals. Weyl quasiparticles have in fact been revealed experimentally, not only in solid-state systems~\cite{Xu15,Lv15}, but also in photonic~\cite{Lu13,Lu15} and phononic~\cite{Xiao20} scenarios, where the quasiparticles are of bosonic nature instead. One of the key features of Weyl semimetals is their topological protection against perturbations. Therefore, these systems, along with the so-called Dirac semimetals, constitute a new paradigm of topological phases without the requirement of a bulk energy gap~\cite{Armitage18}. 

Because of their relevance in the field of topological matter, the characterization of Weyl semimetals proves to be necessary towards their potential use in applications. To this regard, the analysis of how these materials behave in presence of impurities becomes essential. In this context, several studies have been conducted in Weyl, Dirac, nodal loop and triple-component semimetals~\cite{Huang13a,Huang13b,Sun15,Chang15,Mitchell15,Zheng16,Deng17,Deng18,He18,Rusmann18,Lu19,Novelli19,Rancati20,Lee21} and topological insulators~\cite{Liu09,Biswas10,Chen10,Zazunov10,Okada11,Beidenkopf11,Lu11,Black12,Black15,Sessi16,Yang18}, where signatures of the Kondo effect are observed, the appearance of resonances is displayed and the stability of these materials against impurities is discussed. In this work, we will work along the lines of Ref.~\cite{Biswas10}, where the effect of single scalar and magnetic impurities at the surface of a topological insulator is considered. In our work, we will consider a Weyl semimetal instead. 
\revision{Furthermore, while most of the previous works deal with zero-range impurity potentials, we introduce an exactly solvable model using a non-local separable pseudopotential~\cite{Sievert73,Prunele97}. In particular, it is worth mentioning that finite-range pseudo-potentials, such as Yamaguchi's~\cite{Yamaguchi54}, can nicely reproduce electron interaction with screened, local Coulomb potentials~\cite{Lopez02}. In addition, our approach is particularly useful when extending the study to many impurities by applying the coherent potential approximation, which would allows us to obtain closed expressions for the density of states, as we showed recently in the case of non-magnetic impurities at the surface of a topological insulator~\cite{Hernando21}}.
Our results present clear similarities with the two-dimensional case of a topological insulator, such as the absence of gap openings~\cite{Biswas10,Sessi16}, revealing that the phenomena discussed herein is inherent to the cone-like structure and not to the dimensionality of the problem. In contrast to the two-dimensional case, the isotropy of the problem disregards the possibility of observing different scenarios where the magnetic impurities are perpendicular or parallel to the surface. However, it allows us to observe spin textures along the three spatial directions. As we shall show, these are highly non-trivial and anisotropic, with clear winding around the impurity. Additionally, Friedel oscillations are observed in the local density of states (LDOS)~\cite{Zheng16}, which could potentially be experimentally observed by using scanning tunneling microsopy (STM)~\cite{Beidenkopf11,Sessi16,Rusmann18}.  

\section{Model}

We will consider a single-node Weyl semimetal with a finite bandwidth that we will denote by $2\Delta$. Thus, the single particle Hamiltonian for the Weyl semimetal will be
\begin{equation}
    H_0(\bm{k})=\bm{\sigma}\cdot\bm{k} \ ,
    \label{eq:01}
\end{equation}
where energies are measured in units of $\Delta$ and momenta in units of $1/\ell$ with $\ell=\hbar v/\Delta$ and $v$ the Fermi velocity. Here, $\bm{\sigma}=(\sigma_x,\sigma_y,\sigma_z)$ are the Pauli matrices \revision{and corresponds to a pseudospin degree of freedom. It must be noted that, although single node models are suitable when considering continuum descriptions~\cite{Armitage18}, a study of real Weyl materials needs to consider an even number of Weyl fermions with chiralities such that the total chirality adds up to zero. This is also necessary in order to study the topology of Weyl materials, together with the appearance of surface states and Fermi arcs that connect both nodes. Moreover, translational symmetry breaking can affect the robustness of the Weyl nodes~\cite{Armitage18}, as it occurs when considering finite materials~\cite{Baba19}. Therefore, a fuller treatment of the problem at hand, where the impurity naturally breaks translational symmetry, would require considering two nodes so as to study the topological protection. This can be achieved by considering quadratic terms in the Hamiltonian~\cite{Baba19}, which complicate matters slightly. However, for small enough impurity strengths, the Weyl nodes should remain effectively protected and decoupled, as can be understood by considering adiabatic continuity. Therefore, we proceed hereon with the single-node model for simplicity, although it must be carefully remembered that this is a first order approximation to the problem.}

The single impurity will be included using a non-local separable pseudopotential of the form~\cite{Sievert73,Prunele97,Lopez02,Lima08,Hernando21}
\begin{equation}
    \hat{V} = |\omega\rangle\lambda\langle\omega| \ , 
    \label{eq:02}
\end{equation}
where $\omega(\bm{r})=\langle\bm{r}|\omega\rangle$ is the so-called shape function, which shall be specified later, $\lambda=U$ for the scalar impurity and $\lambda=U\bm{S}\cdot\bm{\sigma}$ for the magnetic impurity, with $U$ a real number and $\bm{S}$ a unit vector defining the orientation of the impurity's spin. Hence, the full Hamiltonian will be
\begin{equation}
    \hat{H}=\hat{H}_0+\hat{V} \ ,
    \label{eq:03}
\end{equation}
with $\langle\bm{k}|\hat{H}_0|\bm{q}\rangle=H_0(\bm{k})\delta(\bm{k}-\bm{q})$. In the spirit of Ref.~\cite{Biswas10}, we shall be interested in calculating the following quantities: the spin-unresolved local density of states (LDOS)
\begin{equation}
    \rho(\bm{r},E)=-\frac{1}{\pi}\mathrm{Im}~\mathrm{Tr}\left[\langle\bm{r}|\hat{G}(E)|\bm{r}\rangle\right] \ ,
    \label{eq:04}
\end{equation}
the local density of up/down spins in direction $i$
\begin{equation}
    \rho_i^{\pm}(\bm{r},E)=-\frac{1}{\pi}\mathrm{Im}~\mathrm{Tr}\left[\langle\bm{r}|\hat{G}(E)\left(\frac{1\pm\sigma_i}{2}\right)|\bm{r}\rangle\right] \ ,
    \label{eq:05}
\end{equation}
and the energy-resolved spin density average
\begin{equation}
    \bm{s}(\bm{r},E)=-\frac{1}{\pi}\mathrm{Im}~\mathrm{Tr}\left[\langle\bm{r}|\hat{G}(E)\frac{\bm{\sigma}}{2}|\bm{r}\rangle\right] \ ,
    \label{eq:06}
\end{equation}
where $\hat{G}(z)=(z-\hat{H})^{-1}$ is the retarded Green's function of the system with $z=E+\ii 0^{+}$. If we denote the retarded Green's function associated to $\hat{H}_0$ by $\hat{G}_0$, we may find $\hat{G}$ from
\begin{equation}
    \hat{G}=\hat{G}_0+\hat{G}_0\hat{T}\hat{G_0} \ , 
    \label{eq:07}
\end{equation}
with
\begin{equation}
    \hat{T}=\left(1-\hat{V}\hat{G}_0\right)^{-1}\hat{V} \ ,
    \label{eq:08}
\end{equation}
which can be equivalently written as
\begin{equation}
    \hat{T}=|\omega\rangle W\langle\omega| \ ,
    \label{eq:09}
\end{equation}
with
\begin{equation}
 W = \left[1-\lambda\langle\omega|\hat{G}_0|\omega\rangle\right]^{-1}\lambda \ .
 \label{eq:10}
\end{equation}
In order to calculate the aforementioned quantities [cf. Eqs.~(\ref{eq:04})-(\ref{eq:06})] we need to first calculate $\langle\bm{r}|\hat{G}|\bm{r}\rangle$. To do so, we need to specify a shape function. We will choose it such that $\omega(\bm{k})=\langle\bm{k}|\omega\rangle$ is spherically symmetric, i.e. $\omega(\bm{k})=\omega(k)$ with $k=|\bm{k}|$, and it is short-ranged in coordinate space. Therefore, we can write it as $\omega(\bm{k})=\Theta(k_c-k)$, with $\Theta(x)$ the Heaviside step function and $k_c=1$ the momentum cutoff. Notice that $k_c=1$ is the dimensionless momentum corresponding to $\Delta/\hbar v$. \revision{In what follows, it must be  therefore be remembered that $k_c$ is not a parameter.} Let 
\begin{equation}
    P(E,r) = -\,\frac{1}{\sqrt{2\pi}r}\,\left[\ii\pi\sin(Er)+C(E,r)\sin(Er)+S(E,r)\cos(E,r)\right] \ ,
    \label{eq:11}
\end{equation}
being $C(E,r)$ and $S(E,r)$ combinations of the sine and cosine integral functions
\begin{equation}
    \begin{aligned}
    C(E,r)& =\mathrm{Ci}\left[(k_c-E)r\right]-\mathrm{Ci}\left[(k_c+E)r\right] \ , \\
    S(E,r)& =\mathrm{Si}\left[(k_c+E)r\right]-\mathrm{Si}\left[(k_c-E)r\right] \ .
    \end{aligned}
    \label{eq:13}
\end{equation}
Then, after some algebraic manipulations, \revision{detailed in the Supplementary Material}, we arrive at the following expression for $\langle\bm{r}|\hat{G}|\bm{r}\rangle$
\begin{equation}
    \langle\bm{r}|\hat{G}|\bm{r}\rangle = G_0(E) + Q(E,r)WQ(E,-r) \ , 
    \label{eq:14}
\end{equation}
where
\begin{equation}
    G_0(E)=-\frac{E}{4\pi^2}\left(\ii E\pi+2k_c-E\ln\frac{k_c+E}{k_c-E}\right) \ ,
    \label{eq:15}
\end{equation}
and
\begin{equation}
    Q(r,E)=\alpha(E,r)+\beta(E,r)\sigma_r \ ,
    \label{eq:16}
\end{equation}
with $\alpha(E,r)=EP(E,r)$, $\beta(E,r)=-\ii\partial_rP(E,r)$ and $\sigma_r=\bm{\sigma}\cdot\widehat{\bm{r}}$. \revision{It must be noticed that $G_0(E)$ as it appears in Eq.~\eqref{eq:15} is the unperturbed Green's function at the origin when considering a finite bandwith, as shown by the presence of $k_c$. In any case, the terms dependent on $k_c$ in Eq.~\eqref{eq:14} are irrelevant in the LDOS and related quantities when taking the imaginary part. However, $G_0(E)$ will appear as such, with the cutoff, in the expressions given below for $W$, as shown in the Supplementary Material.
}
\section{Scalar impurity}
In this section we consider the case of a scalar impurity, that is, one where $\lambda=U$ with $U$ a real number. From the previous results, we find that the spin-unresolved LDOS is given by
\begin{equation}
    \rho(r,E) = \frac{E^2}{2\pi^2}-\frac{2}{\pi}\mathrm{Im}\left[\frac{\alpha^2(E,r)-\beta^2(E,r)}{\gamma_s(E)}\right] \ , \label{eq:17}
\end{equation}
where
\begin{equation}
    \gamma_{s}(E)=U^{-1}-(2\pi)^3G_0(E) \ .
    \label{eq:18}
\end{equation}
The symmetry of the problem implies that the spin-resolved LDOS along any direction is simply half of the LDOS, that is,
\begin{equation}
    \rho_i^{\pm}(r,E)=\frac{1}{2}\rho(r,E) \ ,
    \label{eq:19}
\end{equation}
as can also be seen directly from Eq.~(\ref{eq:05}). Finally, the energy-resolved spin density average is identically zero, as can be seen from Eq.~(\ref{eq:06}).

\begin{figure}[htb]
    \centering
    \includegraphics[width=1\textwidth]{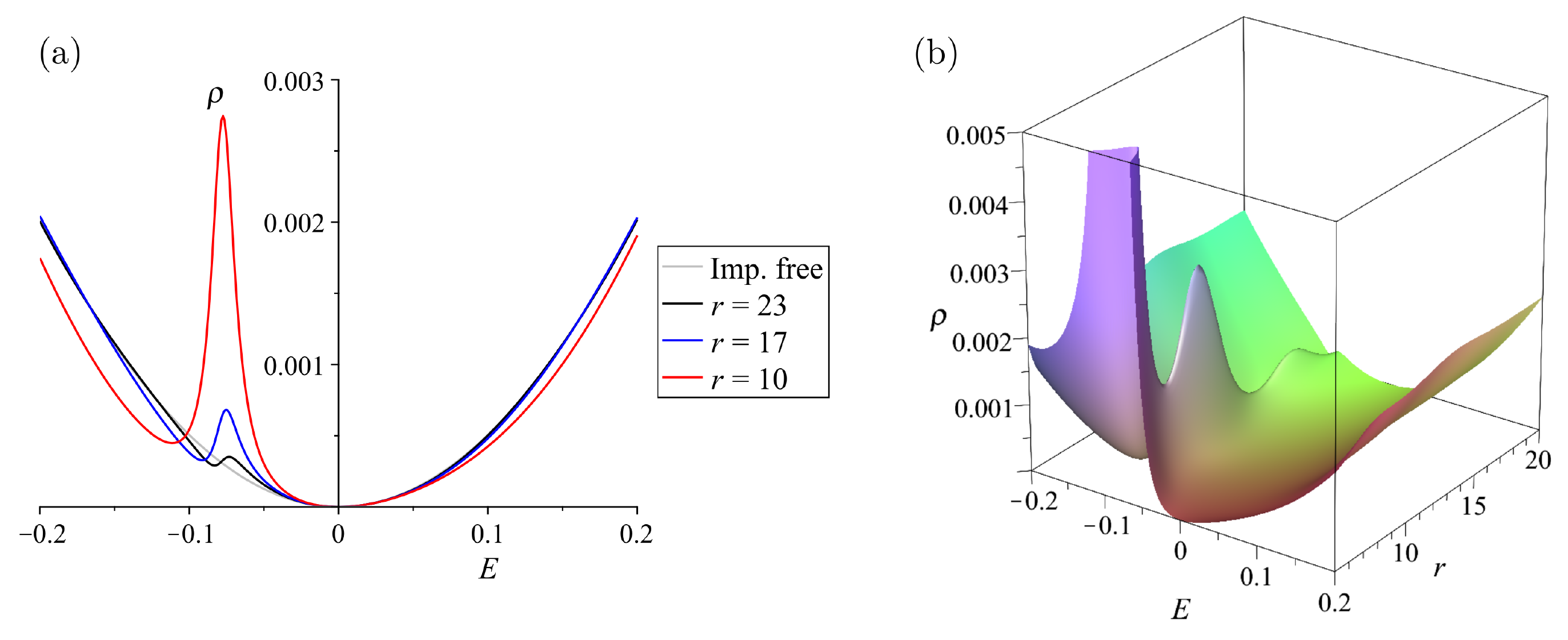}
    \caption{(a)~LDOS as a function of energy for three values of $r=10,17$ and $23$. Also shown is the impurity-free LDOS (gray curve). The peak gets reduced in size as the distance to the impurity increases and it also displaces towards zero energy. (b)~LDOS as a function of energy and distance to the impurity. Friedel oscillations are observed as the LDOS gets reduced in an oscillatory fashion when $r$ increases.}
    \label{fig1}
\end{figure}

In Fig.~\ref{fig1}(a) we show the LDOS for three values of $r$. We will be considering $U=1$ hereafter. In the following section where the magnetic impurity is considered, this value of $U$ will be justified with experimentally feasible values.
As it can be observed in the figure, a resonance peak in the otherwise quadratic LDOS appears, which gets reduced as the distance to the impurity increases. Moreover, we observe that the resonance gets shifted towards zero energy as the distance increases, so that far from the impurity the parabolic LDOS is recovered. The reduction in size of the resonance peak occurs accompanied by Friedel oscillations, as can be observed in Fig.~\ref{fig1}(b). The occurrence of Friedel oscillations can be traced back to the interference of incoming and outcoming waves due to scattering processes at the impurity~\cite{Wang15}. 

\section{Magnetic impurity}

\begin{figure}[htb]
    \centering
    \includegraphics[width=1\textwidth]{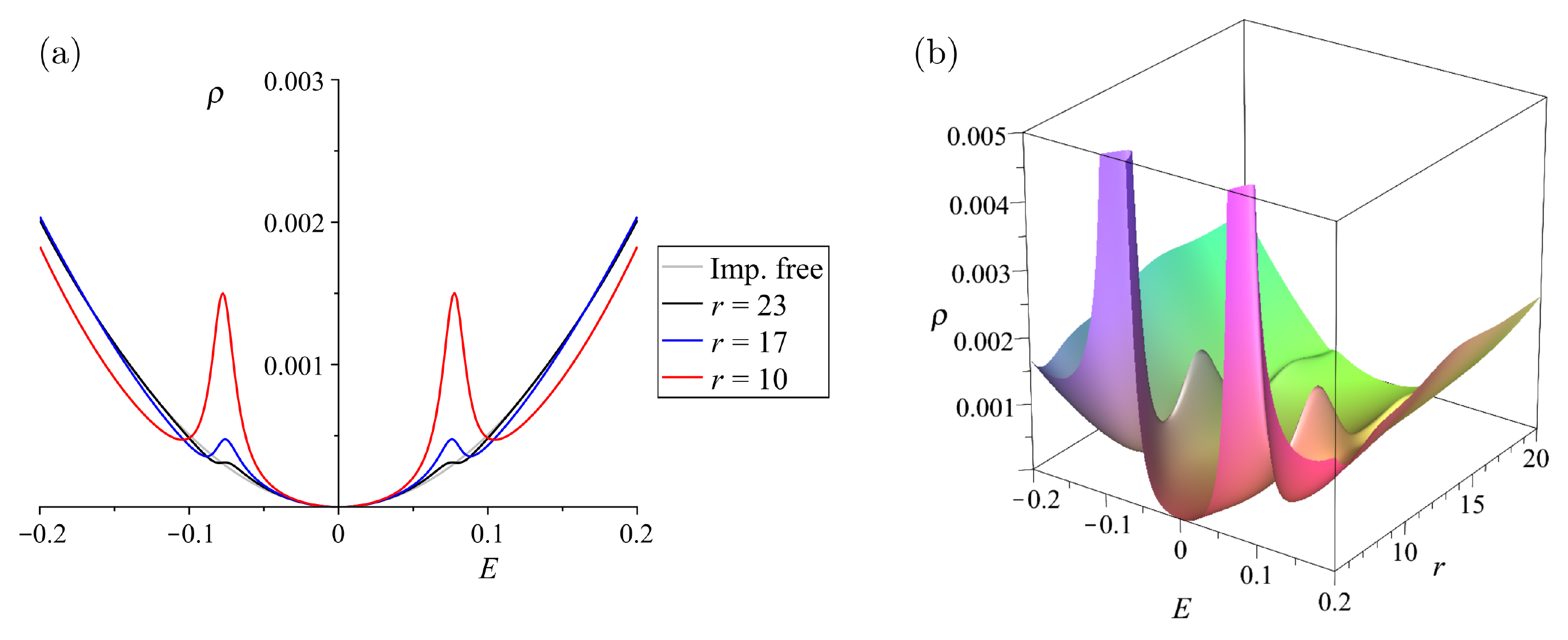}
    \caption{(a) LDOS as a function of energy for three values of $r=10,17$ and $23$ in the case of a magnetic impurity. The peaks follow the same trend as in the scalar case as $r$ increases. The size of the peaks is half that of the scalar case. (b) LDOS as a function of energy and distance to the magnetic impurity. Friedel oscillations are observed as the LDOS gets reduced in an oscillatory fashion as $r$ increases.}
    \label{fig2}
\end{figure}

We now turn our attention to the case of a magnetic impurity. Because of the spherical symmetry of the problem, the spin of the impurity can point along any direction of our choice. Thus, we will choose the spin to point along the $z$-direction, i.e. $\bm{S}=\widehat{\bm{z}}$. Let
\begin{equation}
    \gamma_m(E)=\frac{1}{U^2(2\pi)^3G_0(E)}-(2\pi)^3G_0(E) \ .
    \label{eq:20}
\end{equation}
Then, the LDOS is given by the same expression as Eq.~(\ref{eq:17}) with $\gamma_s(E)\to\gamma_m(E)$. In contrast to the unresolved LDOS, the spin-resolved LDOS depends on $r$ but also on the polar angles $\theta$ and $\varphi$. Indeed, we find that
\begin{equation}
    \begin{aligned}
    \rho_x^{\pm}(r,\theta,\varphi,E)& =\frac{E^2}{4\pi^2}-\frac{1}{\pi}\mathrm{Im}\bigg[\frac{\alpha^2-\beta^2}{\gamma_m}(E)\nonumber\\ 
    &\mp\beta^2\nu(E)\sin(2\theta)\cos\varphi\pm 2\ii\alpha\beta\nu(E)\sin\theta\sin\varphi\bigg] \ , \\
    \rho_z^{\pm}(r,\theta,\varphi,E) & = \frac{E^2}{4\pi^2}-\frac{1}{\pi}
    \mathrm{Im}\bigg[\frac{\alpha^2-\beta^2}{\gamma_m(E)}\pm\alpha^2\nu(E)\mp\beta^2\nu(E)\cos(2\theta)\bigg] \ ,
    \end{aligned} 
    \label{eq:21}
\end{equation}
where we have defined
\begin{equation}
    \nu(E)=\frac{1}{\gamma_m(E)U(2\pi)^3G_0(E)} \ .
    \label{eq:22}
\end{equation}
We have omitted $\rho_y^{\pm}$ since it is obtained from $\rho_x^{\pm}$ by doing $\varphi\to\varphi-\pi/2$. Finally, the spin-resolved density average, as it can be observed from Eq.~(\ref{eq:06}), can be found as
\begin{equation}
    \bm{s}(E,\bm{r})=\frac{\rho_x^{+}-\rho_{x}^{-}}{2}\,\widehat{\bm{x}}+\frac{\rho_y^{+}-\rho_{y}^{-}}{2}\,\widehat{\bm{y}}+\frac{\rho_z^{+}
    -\rho_{z}^{-}}{2}\,\widehat{\bm{z}} \ .
    \label{eq:23}
\end{equation}
Taking into account that the quantity with dimensions corresponding to $U$ is given by $U^{*}=U\Delta\ell^3$ and that $U^{*}=JS/2$, then a value of $U=1$ is reasonable if we consider that typical values for Weyl semimetals of $\hbar v\sim \SI{250}{meV\cdot nm}$~\cite{Arnold16}, the exchange energy $J\sim \SI{300}{meV\cdot nm^3}$~\cite{Matsukura98,Dyck02,Liu14} and a bandwidth cutoff of $\Delta\sim\SI{250}{meV}$. 
In Fig.~\ref{fig2}(a) we show the LDOS as a function of $E$ for the same values of $r$ as those in Fig.~\ref{fig1}(a). As we can see, the single resonance splits into two symmetric resonances with half the size of the one due to the scalar potential, as can be understood from the fact that $\lambda=U$ in the scalar case whereas in the magnetic case we have $\lambda=U\sigma_z$. Similarly, Friedel oscillations appear, as shown in Fig.~\ref{fig2}(b). 

\begin{figure}[htb]
    \centering
    \includegraphics[width=1\textwidth]{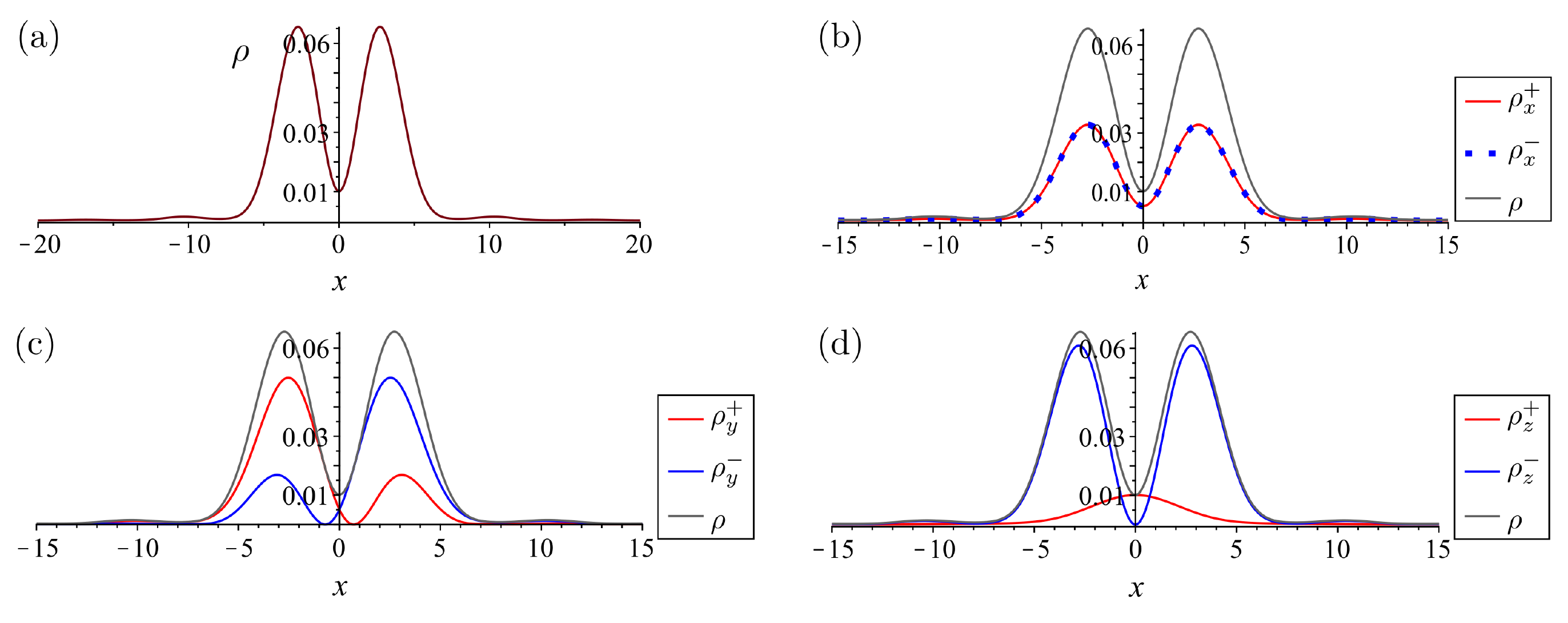}
    \caption{(a) Spin-unresolved LDOS along the $x$-axis at the negative resonance peak. (b), (c) and (d) show the spin-resolved LDOS along the $x$-axis at the negative resonance peak for up/down spins in the $x$, $y$ and $z$ directions, respectively. In (b), the up and down densities coincide, leading to $s_x=0$ along the $x$-axis. In (c), the up and down densities are mirror images of each other. Finally, in (d) there are no down spin states right at the impurity.}
    \label{fig3}
\end{figure}

\begin{figure}[htb]
    \centering
    \includegraphics[width=0.8\textwidth]{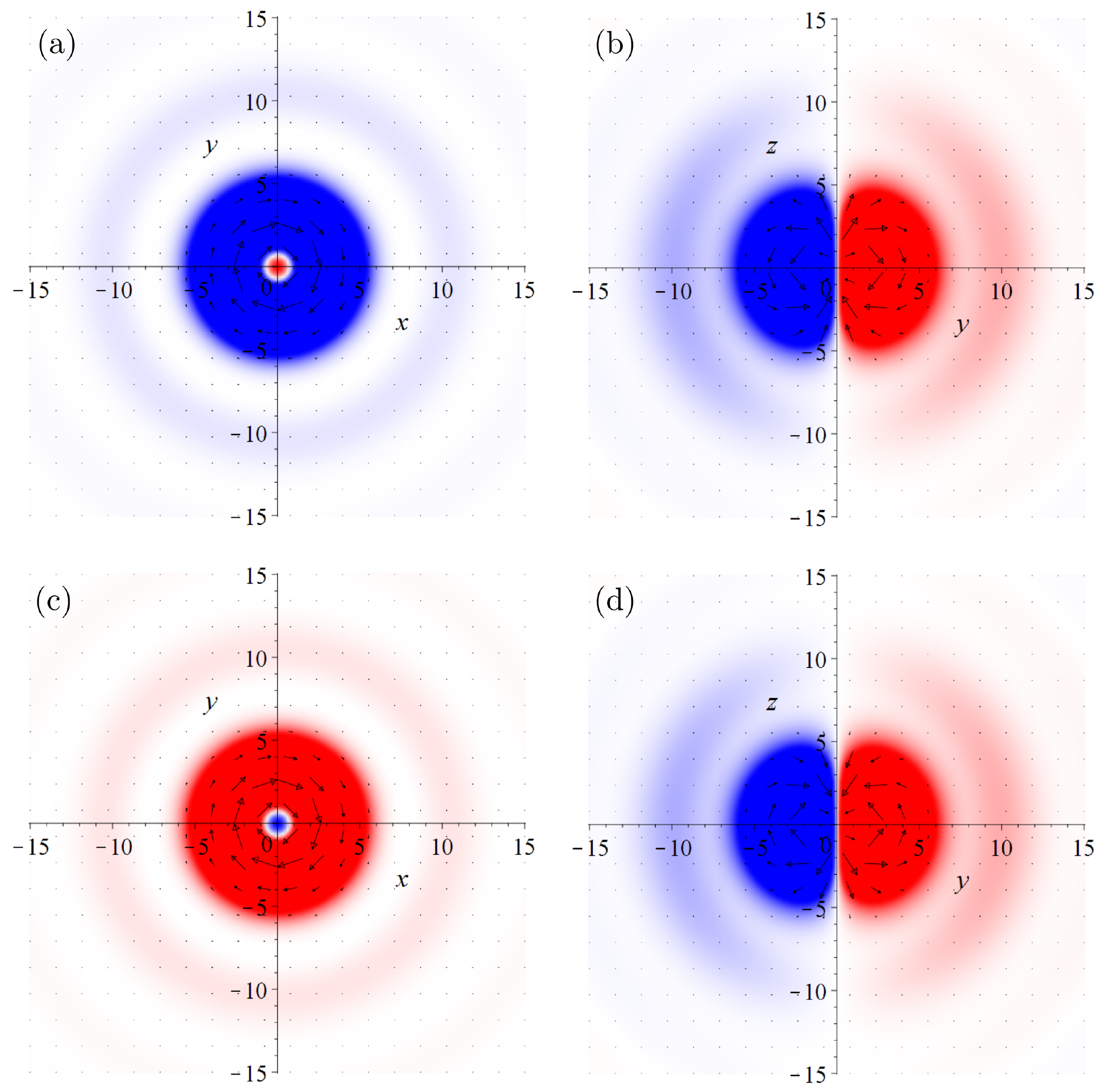}
    \caption{(a) and (b) correspond to the spin textures around the negative resonance in the $xy$ and $yz$ planes, respectively. The same is shown in (c) and (d) for the positive resonance. As it can be seen, the spins circulate in the clockwise direction for both resonances in the $xy$-plane, while having the $z$-component reversed. Friedel oscillations in the LDOS also show up in the spin textures.}
    \label{fig4}
\end{figure}

Next, we show results for the spin-resolved LDOS as a function of distance along the $x$-axis for the energy of the negative resonance peak. In Fig.~\ref{fig3}(a) we show the unresolved LDOS. In Figs.~\ref{fig3}(b) and \ref{fig3}(c), we can see the spin-resolved LDOS in the $x$ and $y$-directions, respectively, as a function of distance $x$ to the impurity. Interestingly, $\rho_{x}^{\pm}$ coincide, whereas $\rho_{y}^{+}$ is the mirror image of $\rho_{y}^{-}$. As we shall see, this means that the spin textures will circulate around the impurity in the $xy$-plane. In Fig.~\ref{fig3}(d), we can see $\rho_z^{\pm}$. As we observe, right at the impurity, which is aligned along the positive $z$-direction, there are no states of down spin along the $z$-direction. Away from the impurity, however, the states become mostly antiparallel to the impurity, similar to what has been observed in the case of topological insulators~\cite{Biswas10}.
Finally, we can observe the spin textures for the negative and positive resonance peaks. The results for the negative resonance are shown in Figs.~\ref{fig4}(a) and (b) and those of the positive resonance are displayed in Figs.~\ref{fig4}(c) and (d). Arrows indicate the components of $\bm{s}$ in the plane shown and colours indicate the size of the projection along the direction perpendicular to the plane, red (blue) colour denoting a dominating component pointing into (out of) the plane. The spin textures reported herein depict non-trivial behaviour, with the components perpendicular to the impurity circulate around it, as it was anticipated previously when observing the spin-resolved LDOS, and similar to what has been seen in the case of topological insulators~\cite{Biswas10}. We can also notice that, although the circulation is in the clockwise direction for both spins, the direction of the components parallel to the impurity are reversed. It is also interesting to observe that the Friedel oscillations discussed earlier show up also in the spin textures, a fact that was not seen to occur in topological insulators~\cite{Biswas10}.

\section{Conclusions}

In the field of topological matter, Weyl semimetals are attracting great interest due to their unusual transport, magnetic and optical properties~\cite{Nagaosa20}. A proper understanding of Weyl semimetals exposed to different perturbations is therefore in order. In this paper, we aim at contributing to such an enterprise by considering the effect of a single magnetic impurity on relevant properties of such systems, such as the LDOS, the spin-resolved LDOS and the spin textures. The analysis presented in our paper constitutes a first step towards a more elaborate study on the presence of multiple random magnetic impurities, which shall be tackled elsewhere. 

In our work, we have shown that certain features displayed in the two-dimensional surface states of topological insulators can also be realised in Weyl semimetals, such as the presence of symmetric resonances around zero energy in the LDOS~\cite{Biswas10}, together with Friedel oscillations occurring due to interference of incoming and outcoming waves from scattering at the impurity. These Friedel oscillations have an effect on the spin textures at the resonance peaks. The spin textures are highly non-trivial and anisotropic, winding around the impurity and becoming parallel to the spin of the impurity right at the location of the latter. Since the values of the impurity strength considered herein are within those for typical Weyl semimetals, we believe that our results could be observed experimentally. In particular, Friedel oscillations could be unraveled by STM measurements~\cite{Beidenkopf11,Sessi16,Rusmann18}. 

\ack

The authors thank Y. Baba for discussions. We acknowledge financial support through Spanish grant PID2019-106820RB-C21.

\section*{References}


\providecommand{\newblock}{}

\end{document}